\documentclass[journal]{IEEEtran}
\usepackage{cite}

\usepackage{xcolor}

\ifCLASSINFOpdf
   \usepackage[pdftex]{graphicx}
\else
\fi
\usepackage{amsmath}
\usepackage{fixltx2e}

\usepackage{stfloats}
\begin{document}
%
\title{Semi-linear all-polarization-maintaining Yb-doped fiber laser oscillator manifesting dual operation regime at net anomalous dispersion}
%
%
%

\author{Mateusz~Pielach,
        Agnieszka~Jamrozik,
        Katarzyna~Krupa,
        and~Yuriy~Stepanenko
\thanks{M. Pielach, A. Jamrozik, K. Krupa, and Y. Stepanenko are with the Institute of Physical Chemistry, Polish Academy of Sciences, Kasprzaka 44/52, 01-224 Warsaw, Poland} \\

\thanks{M. Pielach (mgpielach@ichf.edu.pl), Y. Stepanenko (stepanenko@ichf.edu.pl).}}


%
%

\markboth{arXiv 19.07.2023}%
{Shell \MakeLowercase{\textit{et al.}}: Bare Demo of IEEEtran.cls for IEEE Journals}
%



\maketitle

\begin{abstract}
Ultrafast all-fiber Yb-doped fiber oscillators are usually associated with all-normal-dispersion cavities, which operate in a dissipative soliton regime, quintessential for pulsed operation at the wavelength of 1 \textmu m. This work presents an all-polarization-maintaining Yb-doper fiber laser oscillator that operates in a~dispersion-managed dissipative soliton regime, thanks to incorporating a chirped fiber Bragg grating. The oscillator, mode-locked via a nonlinear optical loop mirror, has an unconventional semi-linear cavity of net anomalous dispersion. Unlike in standard ring resonators, the ultrashort pulse undergoes amplification twice per cavity roundtrip. Additionally, we report a duality of pulsed operation states depending on the pumping power. Strikingly, the oscillator can work in a~subregime similar to the standard dissipative soliton, facilitating further energy scaling at anomalous dispersion. We characterize the low-noise setup capable of delivering pulse energy as high as 6.4~nJ using standard single-mode polarization-maintaining optical fibers.
\end{abstract}


%
\IEEEpeerreviewmaketitle

\section{Introduction}

\IEEEPARstart{U}{}ltrafast Yb-doped fiber laser oscillators entice great scientific interest due to their broad scope of practical applications. They are commonly used as seed sources for more powerful laser systems incorporating several amplification stages. Such fiber-based laser systems can then be further utilized, among others, in micromachining \cite{sugioka2014ultrafast, SZYMBORSKI20211496, stkepak2023high} or nonlinear imaging \cite{NonlinearMicroscopy, Gottschall:12, Brinkmann:19, Marzejon:21}. Thus, fiber lasers can be used not only in research but also in industry and medicine. As a result, they often operate outside of research laboratories in uncontrolled environmental conditions. Consequently, it is indispensable for laser oscillators to be immune to any external factors, such as mechanical vibrations, temperature changes, or high dustiness. Constructing the cavity using only polarization-maintaining (PM) optical fibers solves the problem of vulnerability to the varying external conditions and the issues regarding the oscillator’s instability \cite{Hansel}. Moreover, superior durability and long-term performance can be ensured by exploiting artificial saturable absorbers (SAs), that do not degrade over time \cite{KOBTSEV2022102764}. The most frequently used artificial SAs are: nonlinear optical/amplifying loop mirrors (NOLM\cite{Doran:88, Szczepanek:15}/NALM\cite{Fermann:s, Aguergaray}), Mamyshev oscillators \cite{Regelskis:15, Haig:22}, and cross-splicing nonlinear polarization evolution (NPE) \cite{Szczepanek:18}.

Regardless of the used mode-locking mechanism, all-PM Yb-doped oscillators usually possess all-normal-dispersion (ANDi) cavities \cite{Chong:06, wise2008high}. The design of an all-PM-fiber cavity at 1 micrometer is limited mainly due to the positive group velocity dispersion (GVD) of standard fused silica optical fibers in this spectral region. Despite the lack of standard fibers with anomalous GVD at 1~\textmu m, one can implement specially-designed photonic crystal fibers (PCF) into a cavity \cite{Zhang:13}. Nevertheless, chirped fiber Bragg grating (CFBG) seems more convenient for dispersion management, primarily as the grating can be written in the core of standard passive PM fiber, usually used in the rest of the cavity \cite{woodward2018dispersion}.

Properly arranging the fiber components in a cavity is fundamental when building an optimized all-fiber oscillator. Typically, all-PM-fiber oscillators utilize ring cavities. The most popular structures include \mbox{figure-of-8} \cite{Duling:s} and \mbox{figure-of-9} lasers \cite{Hansel}. Ring resonators are characterized by a single pass through the Yb-doped fiber per cavity roundtrip, whereas the advantage of linear configuration is the double pass through the gain medium. However, when using a design utilizing NOLM or NALM loop without introduced phase shifters, it is impossible to construct an entirely linear cavity, as the SA itself requires the ring part. On the other hand, implementing CFBG in the cavity forces the resonator to have a linear part, as the CFBG is a reflective element. Nonetheless, most setups incorporating a loop mirror and a CFBG are still characterized by a single pass through the gain medium \cite{HybridRingLinear, Ma:21, Pielach:22, Shi:22}.

Over the last few years, plenty of all-PM Yb-doped fiber oscillators utilizing dispersion management with different cavity architectures have been presented. Numerous works compare the operation of the oscillators at different values of net cavity dispersion (NCD). Yu \textit{et al.} \cite{Yu:20} characterized their NPE-based setup for both positive and negative NCD. Our previous work \cite{Pielach:22} also covered this topic in a figure-of-8 NOLM-based laser. Shi \textit{et al.} \cite{Shi:22} have recently demonstrated \mbox{figure-9} all-PM oscillator mode-locked by NALM and characterized it for different values of NCD. Many efforts were also put towards investigating the specifics of the oscillators at near-zero NCD. Ma \textit{et al.}\cite{Ma:21} showed a NALM-based system with a slightly negative NCD of –0.01 ps$^2$. On the other hand, Graf \textit{et al.} \cite{Graf:22} presented a 3x3-NALM-based oscillator with a slightly normal dispersion of 0.01 ps$^2$, that enabled self-similar pulse evolution \cite{PhysRevLett.92.213902}. All these works proved that the operation at near-zero cavity dispersion can reduce the relative intensity noise (RIN) and leads to the broadening of the optical spectrum, paving the way towards delivering shorter pulses than from ANDi systems. Nevertheless, as most of the attention has been paid to near-zero and slightly positive NCD configurations, characterizing oscillators at net anomalous dispersion still constitutes a gap in the state-of-the-art that we fill in this paper. We believe that operation at negative NCD, further away from the zero dispersion, can increase the pulse energy to be comparable to ANDi systems while sustaining a short pulse duration, which might lead to scaling up the peak power directly from the cavity. 

This work demonstrates an ultrafast all-PM Yb-doped fiber laser oscillator that incorporates a CFBG, enabling the operation under a dispersion-managed dissipative soliton regime. We describe an unusual semi-linear cavity that has negative NCD. Furthermore, we investigate two subregimes of operation, a phenomenon which, to the best of our knowledge, has not been observed before at negative NCD. In particular, we focus on a state that manifests characteristics similar to those expected for ANDi setups. Moreover, thanks to the proposed architecture enabling the pulse to undergo amplification twice per cavity roundtrip, we present the largest pulse energy of 6.4~nJ among the all-PM Yb-doped oscillators at~anomalous~NCD.

\section{Experimental setup}

\begin{figure*}[!b] 
\centering
\includegraphics[width=.86\linewidth]{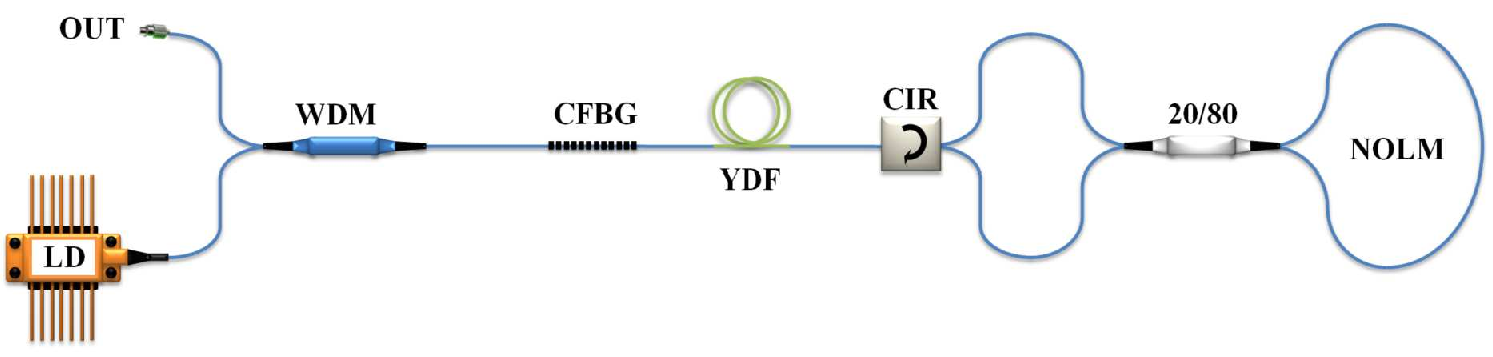}
\caption{Scheme of the semi-linear ultrafast laser oscillator. LD - single mode laser diode (976~nm), WDM - wavelength division multiplexer, CFBG - chirped fiber Bragg grating (-0.237~ps$^2$), YDF - ytterbium-doped fiber (PM-YSF-HI), CIR - optical circulator, 20/80 - 2x2 fiber coupler with a power splitting ratio of 20\% to 80\%, NOLM - nonlinear optical loop mirror, OUT - output port. Passive fiber: PM980-XP.}
\label{Scheme}
\end{figure*}

The semi-linear oscillator is presented in Fig. \ref{Scheme}. The setup was spliced out of standard single-mode, single-clad PANDA type PM optical fibers and fast axis blocked fiberized components. PM980-XP was used as a passive fiber, while a~highly-doped Yb-doped fiber (PM-YSF-HI, Coherent Inc.) was utilized as a gain medium. As a pump, we used a PM-fiber-coupled single-mode semiconductor laser diode operating at 976~nm (LD, 3SP-1999CHP). 0.5 m piece of Yb-doped fiber (YDF) was pumped through a wavelength division multiplexer (WDM), placed outside of the cavity. The used CFBG (DMR-1030-10-25-(+D0.42+0)-P1, TeraXion Inc.) was transmissive for 976~nm and partly reflective for 1030~nm. The reflection spectrum of the CFBG was characterized by the Gaussian shape with full width at half-maximum (FWHM) of 12.8~nm and a maximum reflectivity of 32\% at 1033.1~nm. The CFBG was used as a dispersion-managing element in the cavity, introducing negative group delay dispersion of -0.237 ps$^2$. Considering the GVD of used fibers (0.023 ps$^2$/m), the CFBG could compensate for the second-order dispersion introduced by about 10.3~m of PM980-XP. The reflective grating gave the oscillator a linear cavity part, where we introduced the Yb-doped fiber. This particular architecture permitted the pulse to undergo amplification twice per single cavity roundtrip. The optical circulator (CIR) connected the linear section of the cavity with the ring part, where the artificial SA was placed.

The oscillator was mode-locked via a NOLM build of a~2x2 fiber coupler with a~power splitting ratio of 20/80. Such a~power splitting ratio was optimal in this case, as it provided a steeper increase in the transmission at lower intensities when compared to less asymmetric couplers (e.g.,~30/70) \cite{stkepien2017study, Malfondet:21}. Thanks to that, the length of the NOLM could be shorter ($<$5~m), allowing us to decrease the length of the whole cavity and study the negative NCD configuration. On the other hand, using even more asymmetric couplers (e.g., 10/90) would result in a lower modulation depth and worse background suppression. 

The ports of the NOLM were connected to the circulator in such a way that the NOLM operated as a SA in the transmission mode. The circulator blocked the beam reflected from the NOLM, and the beam transmitted through the NOLM was also transmitted back through the circulator. The pulse transmitted through the circulator propagated through the gain medium and reached the CFBG, where the pulse got partly reflected, compressed, and filtered for the next roundtrip. Due to the limited reflectivity of the CFBG, the majority of the optical power was transmitted through it. The pulse extracted from the cavity reached the WDM, whose port was used as an output.

The laser had no additional bandpass spectral filter, usually present in all-fiber Yb-doped oscillators \cite{grelu2012dissipative}. In our case, the CFBG played the role of a spectral filter. Moreover, an~additional filtering effect was introduced by the NOLM~\cite{Ou:20}.

The NCD could be tuned by changing the lengths of the fibers in the cavity. We focused on the configuration with NCD of (-0.049 $\pm$ 0.002) ps$^2$. Please note that the value of NCD was recalculated based on the repetition frequency measurements. Once we knew the repetition rate and the refractive indices of the used fibers, we could retrieve the information about the total distance over which the pulse propagated in the fiber. Thus, the length of the linear part of the cavity had a double contribution to the final value of NCD.

In the experiments, the optical spectra were measured by an~optical spectrum analyzer (Yokogawa AQ6370C) with a~resolution of 0.1~nm and a sampling of 0.02~nm. We registered the radio frequency (RF) spectra using a 1.2 GHz InGaAs photodiode (DET01CFC, Thorlabs Inc.) and a spectrum analyzer (Agilent Technologies E4443A). 
We used the Ophir 3A thermal power sensor to measure the average output power. Chirped pulse duration was measured by a home-built second harmonic, background-free autocorrelator. 
Moreover, we compressed the output pulses using a compressor employing a pair of parallel diffraction gratings (800 lines/mm) and retrieved the pulse duration using the SPIDER technique.
We observed the pulses with a 6 GHz oscilloscope (Teledyne Lecroy WavePro 7 Zi-A) with a maximum sampling of 40 Gs/s.

\section{Experimental results}

While optimizing the oscillator, we investigated several values of NCD, including positive and negative values, within the range from +0.087 ps$^2$ to -0.092 ps$^2$. We observed the duality of the operation regime for the NCD between \mbox{-0.044~ps$^2$ and -0.071~ps$^2$}. The results below refer to the value of \mbox{(-0.049~$\pm$~0.002)}~ps$^2$, for which we maximized the output pulse energy for one of the subregimes. It is worth noting that for NCD below -0.071~ps$^2$ the operation began to be unstable. We could also observe a noise-like regime.

The duality of operation regime refers to the two states of operation that can be distinguished between each other by, among others, the shape of the optical output spectrum. Typically, it is said that an oscillator with dispersion compensation in a cavity has a~smoother spectral shape than a system with ANDi cavity \cite{Pielach:22, Shi:22}. Here, we observed not only a~previously-known state with smoother spectral slopes, but also a~subregime of the dispersion-managed dissipative solitons with features typical for ANDi systems; namely, the spectrum with specific sharp slopes. We could also observe self-phase-modulation-induced spectral broadening that resulted in the spectral width greater than the CFBG's filter characteristics. In the following, we will denote the subregimes A and B, where state A mimics the characteristics typical for ANDi systems, while state B is a previously-reported subregime with smoother arising slopes of the optical spectrum. 

The mode-locking operation was initiated in state A for the pumping power exceeding 470~mW. Once started, the oscillator could operate in a stable manner after decreasing the pumping power to 460~mW. The laser kept working in state A within the range from 460~mW to 280~mW. While further reducing the pumping power, we could observe a~transition from state A to state B at the pumping power of 280~mW. Surprisingly, when we continued decreasing the pumping power down to 240~mW, there was a transition from state B to state A again. The laser stopped mode-locking for the pumping power lower than 140~mW.

However, it is worth noting that not all the transitions were reversible, and the oscillator manifested the features of a hysteresis. When doing the opposite, that is, after power decrease below 240~mW, we progressively increased the pumping power, and the oscillator switched to state B at 240~mW, as previously. However, at the pumping power of 280~mW, there was no transition to state A anymore. Instead, the oscillator started to operate in subregime B with a continuous wave (CW) component. Further increase in the pumping power led to a~noise-like regime.

\begin{figure}[htbp!]
\centering
\includegraphics[width=0.99\linewidth]{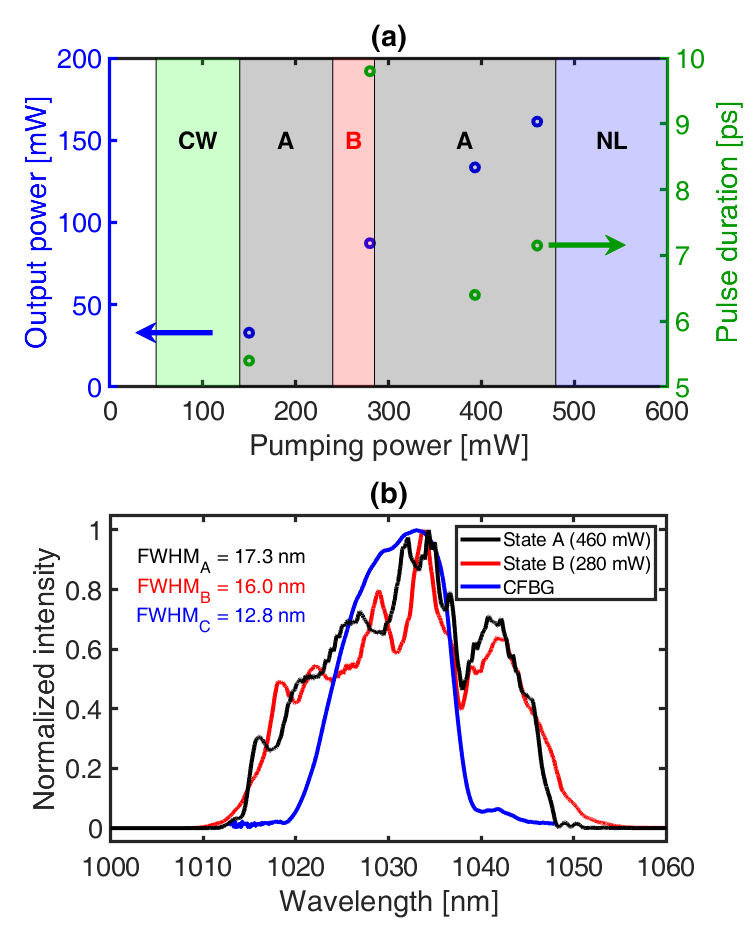}
\caption{Characteristics of the oscillator at NCD of  (-0.049 $\pm$ 0.002) ps$^2$. (a)~Scatter plot of the average output power (blue) and chirped pulse duration directly from the oscillator (green) versus pumping power. CW - continuous wave; A, B - subregimes of operation; NL - noise-like regime. (b) State A (black curve) features sharp slopes in the optical output spectrum, while the spectrum of the subregime B (red curve) arises in a smoother, Gaussian-like manner. The spectral width of both states is greater than the CFBG's reflection spectrum, presented as a blue curve.}
\label{Fig2}
\end{figure}

The map of the subregimes as a function of the pumping power is presented in Fig. \ref{Fig2}(a). We characterized the oscillator in more detail for four measurement points, as the laser was operating stably only in specific ranges of pumping power. Naturally, the higher the pumping power, the higher the average output power (shown as blue circles), which led to higher output energies. The green circles present the results of the autocorrelation measurement directly from the oscillator. Notably, there is a visible increase in the pulse duration for state B, while for state A the chirped pulse duration is getting gradually longer and still maintains shorter than for state B despite greater pumping.
For state A, the higher the pumping power, the broader the optical spectrum due to the self-phase modulation effect, thus longer pulse duration, as more new frequency components were generated. We believe that the sudden increase of the pulse duration for subregime B was due to the different spectral shape of this state. The differences in the spectral characteristics are presented in Fig. \ref{Fig2}(b), which shows the output spectrum for state A (pumping power of 460~mW) in black, and state B (pumping power of 280~mW) in red. Please note that both states A and B were characterized by FWHM values larger than the CFBG's reflection spectrum characteristics, shown as a blue curve in Fig. \ref{Fig2}(b), meaning that they both possessed features of dissipative solitons \cite{grelu2012dissipative}. There were some similarities in the modulations (e.g., for 1040~nm). However, the main difference manifested at the slopes of the spectrum. For state B, we could observe the spectrum arising in a smoother, Gaussian-like manner. Those features have already been reported \cite{Pielach:22}. On the other hand, subregime A was characterized by sharply arising slopes and self-phase-modulation-induced sidelobes. Those features are typical for ANDi systems instead \cite{Chong:06}. To our knowledge, such spectral characteristics have not been observed at anomalous NCD yet.

\begin{figure}[htbp!]
\centering\includegraphics[width=0.91\linewidth]{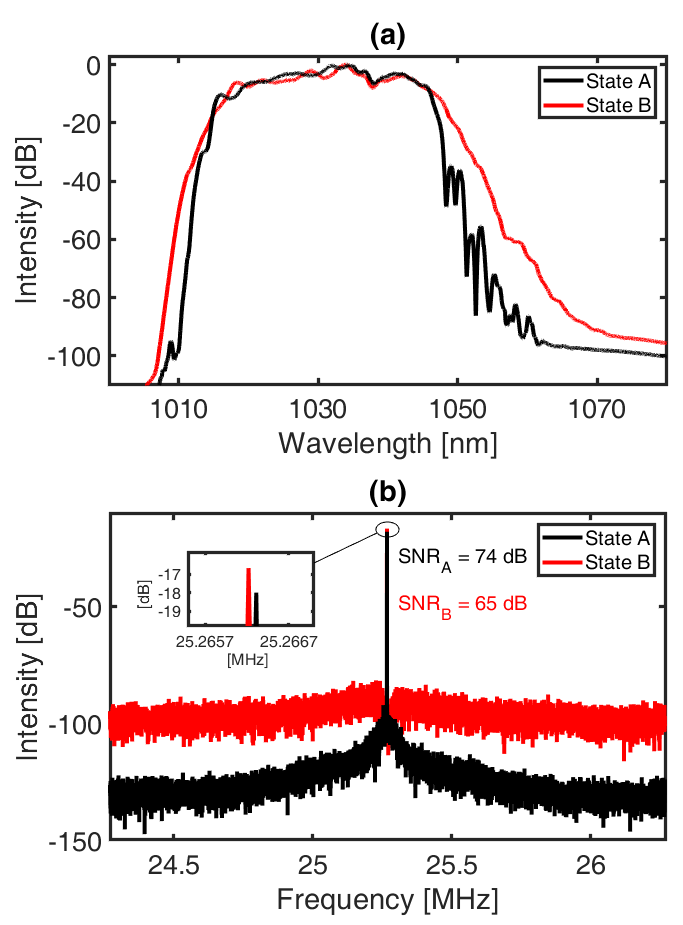}
\caption{Comparison of the characteristics of the oscillator for state A at the pumping power of 460 mW (black curves) to state B at the pumping power of 280 mW (red curves). (a) Optical output spectra in logarithmic scale, proving the different nature of spectral broadening between the two states. (b) The radio frequency spectra indicate a better signa-to-noise ratio for state A, measured in the span of 2 MHz with a resolution bandwidth of~100~Hz.}
\label{Fig3}
\end{figure}

Even though the FWHM value was larger for state A, subregime B was characterized by broader spectra when not considering high-amplitude modulations and considering a~10~dB spectral bandwidth. The comparison of the spectral characteristics in the logarithmic scale is shown in Fig.~\ref{Fig3}(a). We observed that subregime B lacked significant self-phase-modulation-induced sidelobes. The nature of spectral broadening could be seen at the right slope for longer wavelengths. Notably, the differences were not as considerable for shorter wavelengths because the WDM also acted as a longpass spectral filter with a threshold value of around 1005 nm. Moreover, both states were Raman-free, as in neither case did we observe stimulated Raman scattering components, which usually disturb the pulsed operation \cite{Aguergaray:13, Szczepanek:s}.

The oscillator operated at the repetition rate of (25.266~$\pm$~0.001)~MHz, resulting from the cavity's length. Fig.~\ref{Fig3}(b) compares the obtained RF spectra for state A at the pumping power of 460 mW, presented in the black curve, to state B at the pumping power of 280~mW, shown in the red curve. The RF spectra were recorded with a~resolution bandwidth of 100 Hz in a span of 2 MHz. A single peak in the RF spectra proved that both states operated in a~single-pulse regime. However, we observed that subregime A was more stable than state B, as confirmed by a 9~dB higher signal-to-noise ratio (SNR) for state A (74 dB SNR) when compared to state B (65~dB SNR).

\begin{figure*}[htbp!]
\centering\includegraphics[width=\linewidth]{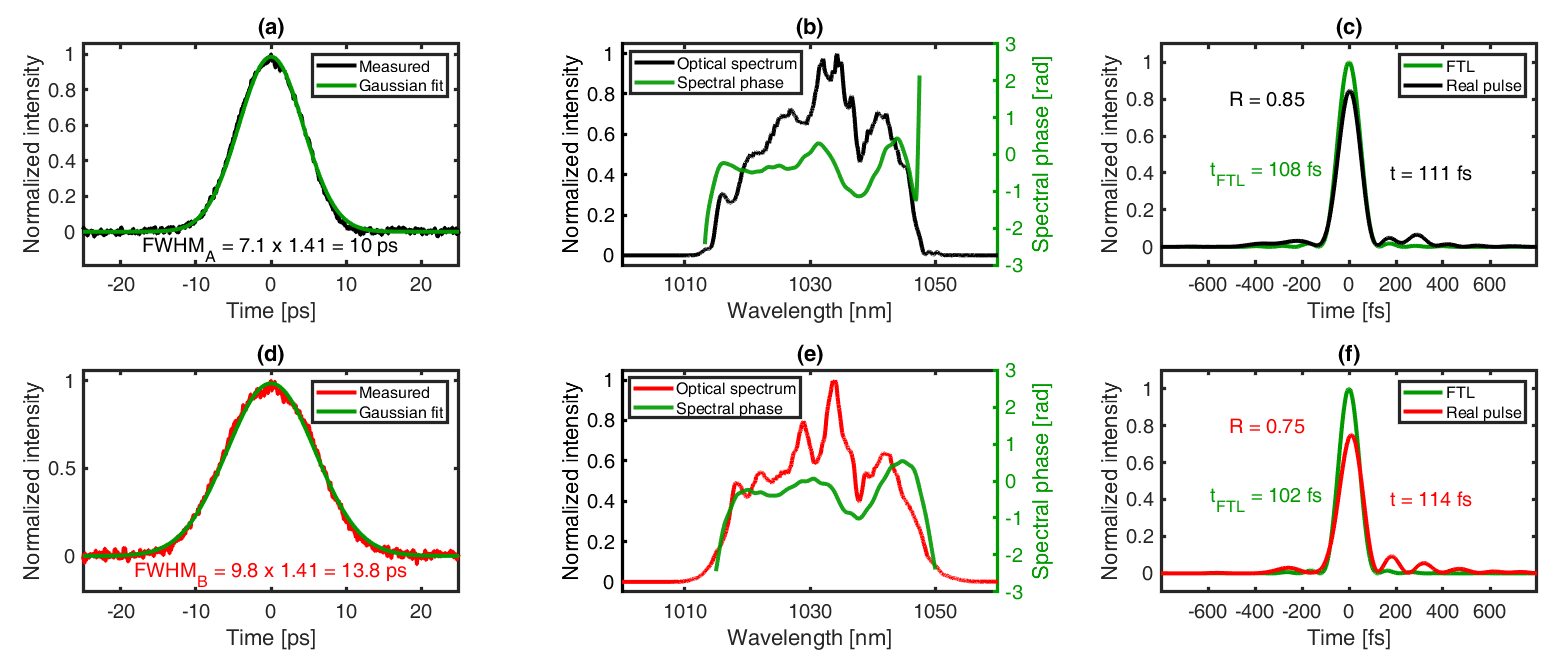}
\caption{Comparison of the temporal characteristics of the oscillator for state A at the pumping power of 460~mW (upper row: a, b, c) to state B at the pumping power of 280~mW (bottom row: d, e, f). (a, d) Autocorrelation traces of the chirped pulses for state A (a) indicating pulse duration of 7.1~ps, and 9.8~ps for state B (d). (b, e) Optical spectrum (black for state A, red for state B) and spectral phase (green), indicating minor higher-order components limiting the pulse duration for state A (b) and state B (e). (c, f) Reconstructed temporal electrical field intensity of the compressed pulse (black for state A, red for state B) compared to Fourier-transform-limited (FTL) pulse duration (green) for state A (c) and state B (f). R - temporal Strehl ratio.}
\label{Fig4}
\end{figure*}

The comparison of the two subregimes for the highest pumping powers in the time domain is presented in Fig. \ref{Fig4}, where the upper row refers to state A, and the bottom row to state B. The pulses directly from the oscillator were positively chirped. Firstly, the pulses were chirped because once they had left the cavity, they propagated for some distance in a fiber with a positive GVD. Secondly, the CFBG played a role of a compressor only for pulses that were reflected from the CFBG. The chirp of the transmitted pulse remained uncompensated. The positive chirp is another feature typical for dissipative solitons. The autocorrelation traces for states A and B are shown in Fig. \ref{Fig4}(a) and \ref{Fig4}(d), respectively. The duration of the pulse directly from the oscillator for state A was (7.1~$\pm$~0.1)~ps, while for state B it elongated up to (9.8~$\pm$~0.1)~ps. We believe that extracting shorter pulses from the semi-linear cavity would be possible by introducing an output coupler inside the cavity, so that pulses could be extracted just after getting reflected from the CFBG. However, due to the limited reflectivity of the CFBG, it was the most efficient to use the beam transmitted through the CFBG as an output port. Nevertheless, the chirped pulse duration was slightly shorter when compared to ANDi systems, which for spectral bandwidths exceeding 15 nm often deliver pulses longer than 10~ps \cite{Aguergaray, Szczepanek:15, Ou:20, Pielach:20, BORODKIN2021107353, Guo:22}.

We compressed the output pulses with a free-space optical gratings compressor to remove the linear chirp. The compressed pulses were characterized using the SPIDER technique \cite{Iaconis:98}. Fig. \ref{Fig4}(b) and \ref{Fig4}(d) present the measured spectral phase in green and the corresponding optical spectra for states A and B, respectively. Despite slight higher-order components in the spectral phase, the pulses could be compressed close to their Fourier transform limits (FTL), presented as green curves in the reconstructed temporal traces in Fig. \ref{Fig4}(c) and \ref{Fig4}(e). Indeed, the temporal Strehl ratio (R) proved the high quality of the compressed pulses. For state A, we obtained (111~$\pm$~3) fs and a temporal Strehl ratio of 0.85, while the FTL duration was 108 fs. On the other hand, for state B, we compressed the pulse to (114 $\pm$ 3) fs with a slightly worse temporal Strehl ratio of 0.75. Nevertheless, the result was close to the FTL \mbox{of~102~fs}.

\begin{figure}[htbp!]
\centering\includegraphics[width=0.99\linewidth]{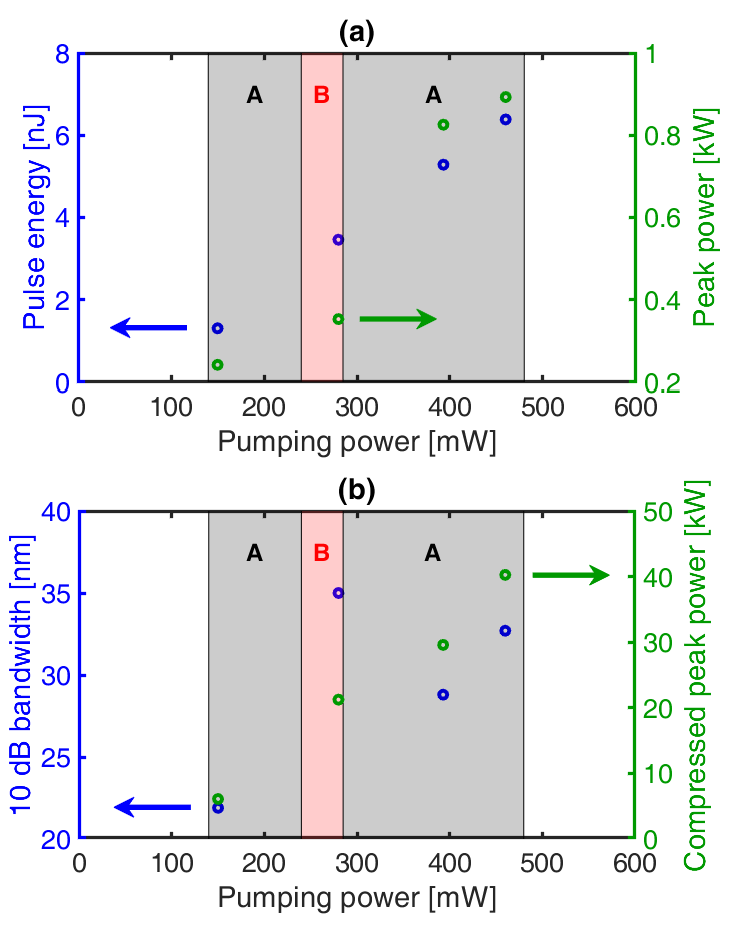}
\caption{Comparison of the characteristics for different pumping powers. (a)~Scatter plot of the output pulse energy (blue) and peak power directly from the oscillator (green) as a function of the pumping power. (b) A plot of the 10~dB spectral bandwidth (blue) as a function of the pumping power indicating the broadest spectrum for state B. The green points indicate the peak power values of the compressed pulses.}
\label{FigCon}
\end{figure}

The maximum average output power for state A was observed for the pumping power of 460~mW and was equal to (161.3 $\pm$ 4.8) mW, which corresponded to the pulse energy of (6.38~$\pm$~0.20)~nJ. For subregime B, we obtained the maximum average power of (87.4 $\pm$ 2.6) mW, corresponding to the pulse energy of (3.46 $\pm$ 0.11) nJ. Indeed, these values can be considered as high. The output power and pulse energy are comparable to ANDi oscillators \cite{Aguergaray, Szczepanek:15, BORODKIN2021107353}. We believe pulse energy can be further scaled using large mode area fibers \cite{Pielach:20, Guo:22}. Nevertheless, to the best of our knowledge, 6.38 nJ is the highest pulse energy among all-PM Yb-doped ultrafast oscillators at negative NCD \cite{Pielach:22, KOBTSEV2022102764}. A more detailed analysis of the dependence of the pumping power on the output pulse energy is presented in Fig. \ref{FigCon}(a). The green circles correspond to the peak power for uncompressed pulses directly from the oscillator. The higher the pumping power, the higher the output energy and the peak power, which reached 0.9 kW for state~A.

The peak power could be boosted by compressing the pulses. When considering the limited diffraction efficiency of the used compressor, the maximum peak power after compression was (40 $\pm$ 1) kW for state A and (21.2 $\pm$ 0.5) kW for state B. The influence of the pumping power on the peak power for compressed pulses is depicted in Fig. \ref{FigCon}(b) in green. In all cases, the output pulses could be compressed very close to the FTL pulse duration, leading to a monotonic increase of the peak power for compressed pulses for higher pumping power. However, it is worth noting that the FTL was changing with the change of the pumping power. The reason is the change in the spectral bandwidth, which is presented in blue in Fig \ref{FigCon}(b). As we discussed before, the 10~dB spectral width was the highest for state B. The change in the spectral bandwidth for state A was caused by the self-phase-modulation-induced spectral broadening. The higher the pumping power, the more substantial the impact of the self-phase modulation, thus the broader the optical spectrum.

\begin{figure*}[htbp!]
\centering\includegraphics[width=0.99\linewidth]{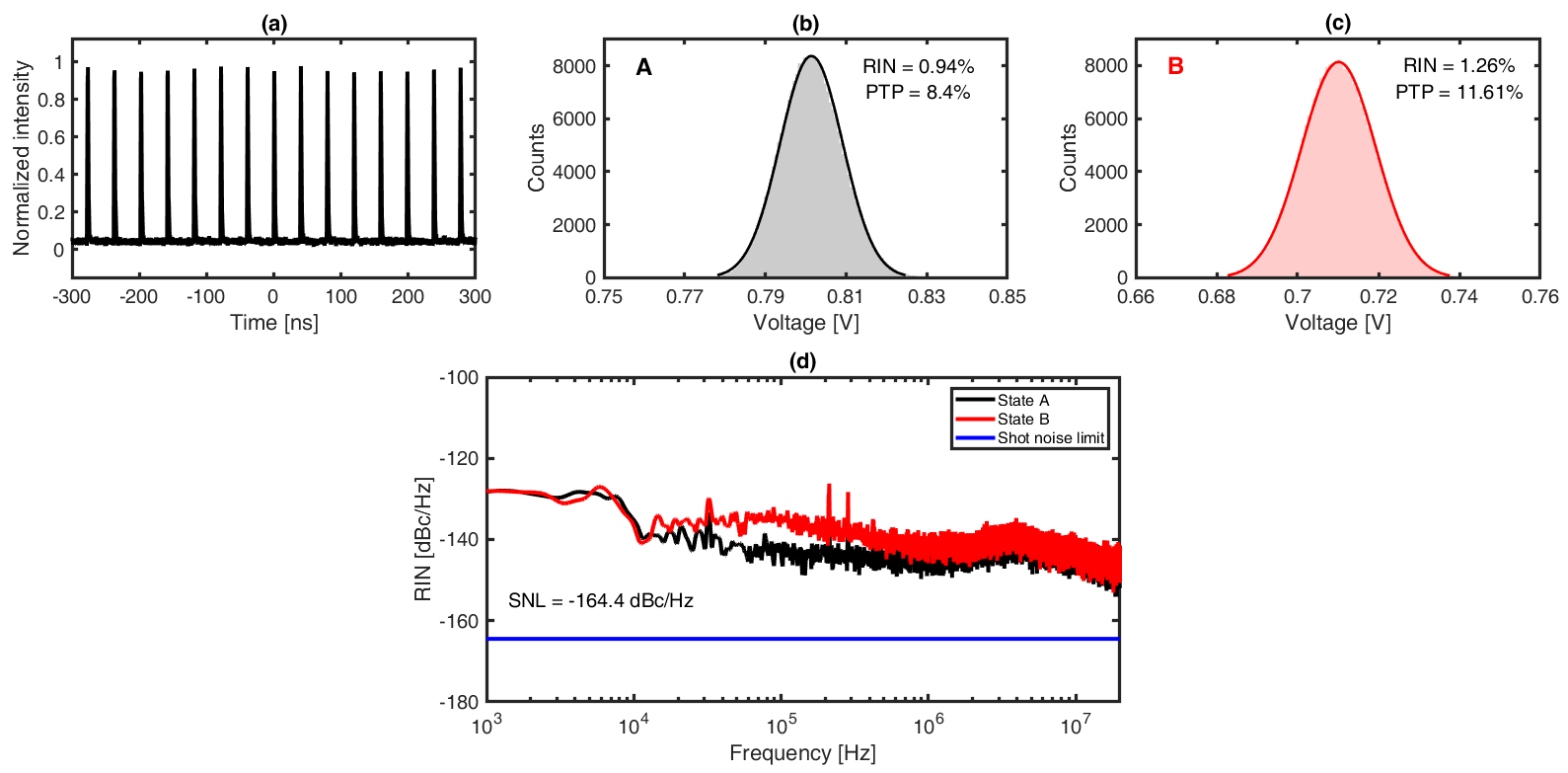}
\caption{RIN measurements in the time domain (upper row) and frequency domain (bottom row). (a) Zoom on an exemplary output pulse train. (b) Histogram of amplitude values for state A at the pumping power of 460 mW indicating RIN value of 0.94\% and peak-to-peak value of 8.4\%. The Gaussian fit is marked in black. (c) Histogram of amplitude values for state B at the pumping power of 280 mW resulting in RIN value of 1.26\% and peak-to-peak value of 11.61\%. The Gaussian fit is marked in red. (d) Representation of the relative intensity noise in the frequency domain for state A (black) and state B (red). SNL - shot noise limit.}
\label{Fig5}
\end{figure*}

We measured RIN to compare the noise characteristics between the two states. First, we estimated pulse-to-pulse RIN in the time domain, following the measurement procedure described by Smith \textit{et al.}\cite{Smith:22}. We used a 1.2 GHz InGaAs photodiode (Thorlabs DET01CFC) that guaranteed a rise time of around 1 ns, which was an order of magnitude shorter than the period of the pulses (40 ns) and longer than the pulse duration (up to 10 ps). We registered ten traces, each in a~total time of 1 ms with a sampling rate of 40 Gs/s, resulting in a~total of about 250k pulses. Fig. \ref{Fig5}(a) presents a~part of an exemplary output pulse train. The dataset for each state consisted of 10 pulse trains, which were registered with a~random time interval between each of the trains, meaning that 250k pulses were not consecutive. Then, we analyzed the amplitude fluctuations of the recorded pulses. The results in the form of a histogram are presented in Fig. \ref{Fig5}(b) and \ref{Fig5}(c) for states A and B, respectively. Note that some of the fluctuations could be caused by the undersampling of the signal, leading to an increase in the RIN value. Nevertheless, this analysis was enough to prove better stability of state A, which was characterized by a RIN value of 0.94\% and peak-to-peak (PTP) value of 8.4\%. We registered a RIN of 1.26\% and a PTP value of 11.61\% for state B. In both cases, the obtained histograms conformed to a Gaussian distribution. We believe that the primary source of the noise was the white noise from the system electronics and the background.

The representation of the RIN in the frequency domain was obtained by calculating the power spectral density through the periodogram of the measurements in the time domain \cite{Quinlan:13, doi:10.1063/1.5129212}. We utilized windowed Fourier transform with Hann window and additionally zero-padded the input signal to increase the sampling in the Fourier domain to 200 Hz. The results in the frequency range from 1 kHz to 20 MHz are presented in Fig. \ref{Fig5}(d). The representation in the frequency domain agreed with the temporal measurements, indicating that the RIN of state A was lower than the RIN of state B. At 1 MHz RIN was determined to be (-144 $\pm$ 2) dBc/Hz for state A and (-141 $\pm$ 2) dBc/Hz for state B. In both cases, the noise stayed below -125 dBc/Hz for the whole measurement range. Please note that these RIN values were at least 5 dBc/Hz higher when compared to an all-PM NALM-based oscillator with NCD closer to zero \cite{Ma:21}. 

\section{Conclusion}
To summarize, we presented the ultrafast all-PM Yb-doped fiber laser oscillator utilizing a semi-linear cavity configuration of anomalous NCD that operates in a dispersion-managed dissipative soliton regime. We observed the duality of pulsed operation states and compared their performance. In particular, we presented a subregime manifesting characteristics similar to ANDi oscillators (state A), that is, to a standard dissipative soliton. This subregime was more stable and delivered shorter pulses of higher energies than state B, a previously reported state at anomalous NCD. The summary of the measured parameters is presented in Tab. \ref{table}. Indeed, the obtained pulse energy and durations are comparable to ANDi systems built out of the same type of PM fiber. Pulse energy exceeding 6 nJ is the highest value among all-PM Yb-doped fiber oscillators utilizing CFBG and operating at negative NCD. 

\begin{table}[!htbp]
 \centering \caption{Comparison of the parameters for two states.}\label{table}
\begin{tabular}{|c|c|c|}
    \hline
    \textit{Parameter} & \textbf{State A} & \textcolor{red}{\textbf{State B}} \\
    \hline
    Pumping power [mW] & 460 & 280 \\
   \hline
    Output power [mW] & 161.3 $\pm$ 4.8 & 87.4 $\pm$ 2.6 \\
   \hline
    Pulse energy [nJ] & 6.38 $\pm$ 0.20 & 3.46 $\pm$ 0.11 \\
    \hline
    Pulse duration [ps] & 7.1 $\pm$ 0.1 & 9.8 $\pm$ 0.1 \\
    \hline
    Peak power from the oscillator [kW] & 0.89 $\pm$ 0.01 & 0.35 $\pm$ 0.01 \\
    \hline
    Transform-limited pulse duration [fs] & 108  $\pm$ 1 & 102 $\pm$ 1 \\
    \hline
    Compressed pulse duration [fs] & 111 $\pm$ 3 & 114 $\pm$ 3 \\
    \hline
    Temporal Strehl ratio & 0.85 $\pm$ 0.01& 0.75 $\pm$ 0.01\\
    \hline
     Peak power of compressed pulses [kW] & 40 $\pm$ 1 & 21.2 $\pm$ 0.5 \\
    \hline
    Pulse-to-pulse RIN [\%] & 0.94 & 1.26 \\
    \hline
   \end{tabular}
    \end{table}

The dual operation regime phenomenon has been most likely caused by the spectral filtering effects of the NOLM~\cite{Ou:20}. Namely, the spectral capabilities of NOLM depend on the energy of the injected pulses. Thus, the spectral width and shape of the NOLM’s filter depend on the pumping power. In our case, NOLM’s spectral bandwidth decreases for higher powers \cite{Ou:20}. Furthermore, the transmission of NOLM increases at larger pumping powers \cite{stkepien2017study, Malfondet:21}. Energy-dependent filtering and transmission of NOLM affect the spectral shape of the pulse propagating in the cavity, which leads to the observation of another pulsed regime \cite{LASZCZYCH2022108107, Wang:17}. Namely, a~shorter pulse of higher energy undergoes overdriven nonlinear effects, inducing the regime transition \cite{xia4499819transient}.

Our research can provide hints for optimizing the Yb-doped fiber laser oscillators operating at anomalous NCD. To conclude, optimizing the oscillator's operation towards the spectra with sharp, steep slopes of the optical output spectra is beneficial. Such ANDi-like characteristics may lead to a~better performance than previously reported subregime with smoother slopes of the optical spectra \cite{Pielach:22}. 
On the other hand, a manifestation of two subregimes with different characteristics may pave the way towards tuning the parameters of the ultrafast fiber oscillators by changing the pump power. Further investigation of the transition between the two states could reveal the ultrafast dynamics behind this phenomenon. We also reckon that the proposed semi-linear configuration might be beneficial in energy scaling. Thanks to a simple cavity configuration and exceptional performance, this oscillator can be used as a seed source for more powerful laser systems.

\section*{Acknowledgment}
This work was funded by the Foundation for Polish Science, under research project TEAM-NET No. POIR.04.04.00-00-16ED/18.

\ifCLASSOPTIONcaptionsoff
  \newpage
\fi



\bibliographystyle{IEEEtran}
\bibliography{IEEEabrv, IEEE_bibfile}
\end{document}